\documentclass[prl,twocolumn,showpacs,preprintnumbers,amsmath,amssymb]{revtex4}
\pdfoutput=1

\usepackage{graphicx}
\usepackage{dcolumn}
\usepackage{bm}
\usepackage[pdftex, bookmarks={false}, pdfauthor={Rudro Rana Biswas}, pdftitle={Notes ..}]{hyperref}
\hypersetup{colorlinks=false, linkcolor=red, citecolor=green, filecolor=blue, urlcolor=blue, filebordercolor={.8 .8 1}, urlbordercolor={.8 .8 0}}
\usepackage[all]{hypcap}

\newcommand\ba{\begin{array}}
\newcommand\ea{\end{array}}
\newcommand\bp{\begin{picture}}
\newcommand\ep{\end{picture}}
\newcommand\be{\begin{equation}}
\newcommand\ee{\end{equation}}
\newcommand\bs{\begin{subequations}}
\newcommand\es{\end{subequations}}
\newcommand\nn{\nonumber}
\newcommand\bfl{\begin{flushleft}}
\newcommand\efl{\end{flushleft}}
\newcommand\bsp{\begin{split}}
\newcommand\easp{\end{split}}

\newcommand\ri{\right}
\renewcommand\le{\left}

\newcommand{\sub}[1]{\mbox{\tiny{#1}}}

\newcommand{\tto}{\rightarrow}
\newcommand{\tr}{\mbox{Tr}}
\newcommand{\sgn}{\mbox{sgn}}

\renewcommand\c{\psi}

\renewcommand\d{\delta}
\newcommand\D{\Delta}

\newcommand\e{\epsilon}

\newcommand\f{\phi}

\newcommand\g{\gamma}

\newcommand\G{\Gamma}

\newcommand\p{\pi}

\newcommand\rr{\rho}

\newcommand\s{\sigma}

\renewcommand\th{\theta}

\newcommand\w{\omega}

\newcommand\x{\xi}
\newcommand\vx{\chi}

\newcommand\y{\eta}

\newcommand\imply{\Rightarrow}

\newcommand\mc{\mathcal}
\newcommand\mb{\mathbb}
\newcommand\mbs{\boldsymbol}

\begin{document}

\title{Quasiparticle Interference and Landau Level Spectroscopy in Graphene\\ in the presence of a Strong Magnetic Field}
\author{Rudro R. Biswas$^{1,3}$}
\email{rrbiswas@physics.harvard.edu}
\author{Alexander Balatsky$^{2,3}$}%
\affiliation{%
$^{1}$Department of Physics, Harvard University, Cambridge, MA 02138\\
$^{2}$Theoretical Division, Los Alamos National Laboratory, Los Alamos, NM 87545\\
$^{3}$Center for Integrated Nanotechnologies, Los Alamos National Laboratory, Los Alamos, NM 87545
}
\date{\today}
\begin{abstract}
We present a calculation of the modulation in the Local Density Of electronic States (LDOS) caused by an impurity in graphene in the presence of external magnetic field. We focus on the spatial Fourier Transform (FT) of this modulation around the impurity. The FT due to the low energy quasiparticles are found to be nonzero over the reciprocal lattice corresponding to graphene. At these lattice spots the FT exhibits well-defined features at wavevectors that are multiples of the inverse cyclotron orbit diameter (see Figure~\ref{fig-results}) and is cut off at the wavevector corresponding to the energy of observation. Scanning Tunneling Spectroscopy (STS) on graphene and the energy-resolved FT fingerprint obtained therefrom may be used to observe the quasiparticle interference of Dirac particles in graphene in the presence of magnetic field.
\end{abstract}
\maketitle

Graphene is a monatomic layer of carbon atoms arranged in a hexagonal lattice. It was first isolated by the mechanical exfoliation of graphite in 2004\cite{2004-novoselov-dc}. The electronic band structure of graphene is characterized by two points $K$ and $K^{*}$ (at wavevectors $\pm\mbs{K}$)\cite{2007-manes-yq} in the reciprocal space where the valence and conduction bands touch each other. The gapless low energy excitations that exist at those points can be described by theories of massless Dirac quasiparticles with opposite chirality in (2+1) dimensions \cite{1947-wallace-yq}. An interesting consequence of such a band structure is the formation of Landau levels (in a perpendicular magnetic field) whose energies vary as the square root of the Landau level index as well as that of the magnitude of the perpendicular magnetic field \cite{1956-mcclure-dn,2009-neto-ys}. The unconventional Quantum Hall Effect seen in transport measurements in graphene is another profound physical consequence of the Dirac nature of these quasiparticles\cite{2005-zhang-mn,2005-novoselov-jt}. Till date, the only evidence of the Landau quantization of Dirac particles has come from bulk transport measurements. Alternately, one might look at the evidence from Landau level spectroscopy using the Scanning Tunneling Microscope (STM). In addition to the real space imaging of Landau levels local spectroscopic tools can be a very sensitive probe of QuasiParticle Interference (QPI) that often reveals details about the underlying band structure and the quasiparticle wavefunctions\cite{2006-balatsky-yq,1997-sprunger-zp}. Applications of these ideas to  graphene are natural and promising. Experiments are currently in progress that probe the signatures of QPI in graphene in the presence of a magnetic field. In this paper we focus on QPI in magnetic field in graphene. We use the theory of non-interacting Dirac quasiparticles in a magnetic field and calculate the change in the electronic LDOS in response to weak impurities. We find that i) the LDOS FT displays characteristic rings whose size is set by the inverse cyclotron diameter $d_{cyc}$ -- indeed, in the limit of a strong magnetic field when effects of disorder and line broadening are secondary the main feature of quasiparticle motion will be the cyclotron orbits -- and ii) these rings will form a lattice in Fourier space that is the same as the graphene reciprocal lattice. Also, depending on the detailed impurity potential structure this FT could have additional angular dependence in $k$-space determined by off-diagonal (sublattice mixing terms) in impurity scattering matrix. These ring-like signatures could be observed in STM experiments.

\begin{figure}
\resizebox{8cm}{!}{\includegraphics{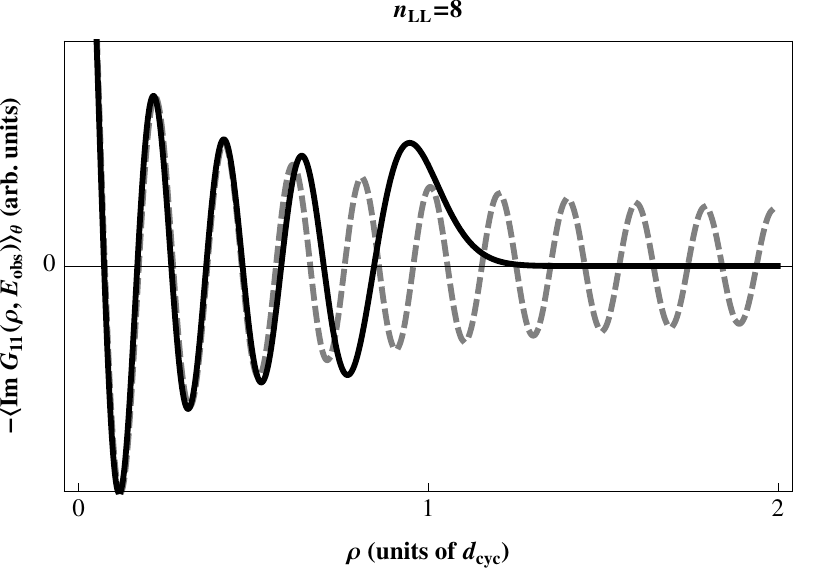}}
\caption{Comparing the Green's functions when $B=0$ (dashed gray) and $B\neq 0$ (continuous black), at an energy $E_{\sub{obs}}$ corresponding to the Landau level index $n_{LL}=8$. The distance propagated is measured in units of the classical cyclotron orbit diameter $d_{cyc} = 2|E|/(evB) = 2\sqrt{2n_{LL}}\ell_{B}$. At small distances these oscillate together at the wavevector $E_{\sub{obs}}/(\hbar v)$ but after $n_{LL}/2$ oscillations the green's function for $B\neq 0$ decays exponentially since the particle `turns' in its cyclotron orbit and cannot propagate further than $d_{cyc}$.}
\label{fig-comparegfs}
\end{figure}
\begin{figure*}
\resizebox{16cm}{!}{\includegraphics{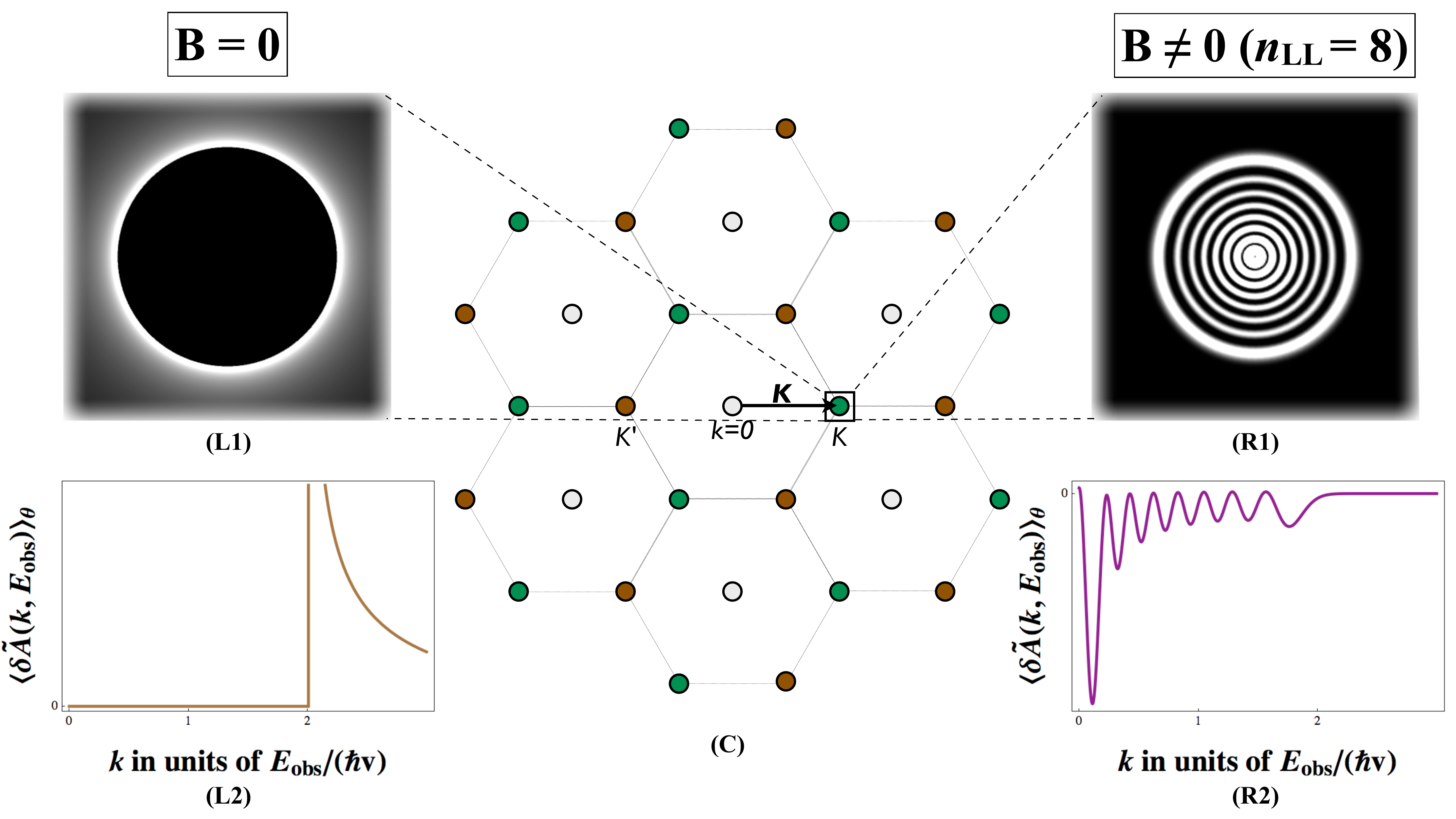}}
\caption{Comparing the \emph{angular averages} of the spatial FT and power spectrum of LDOS modulations around a short-ranged impurity potential $\tilde{\mb{V}}(\mbs{k}) \propto \mb{I}$, for the cases when the magnetic field is zero and when it's nonzero (and the nearest Landau level has an index $n_{LL} = 8$). The center figure (C) shows part of the reciprocal lattice formed by regions in $k$-space where the Fourier transform may be nonzero. The green and red `spots' arise from $K\tto K'$ scattering and vice versa respectively. The grey spots arise from intravalley scattering. One green region is enlarged to show the angle-averaged \emph{Power Spectrum} on a scale where the oscillations are better resolved (the density maps have edges of length $6E_{\sub{obs}}/(\hbar v)$), for the $B\neq0$ (R1) and $B=0$ (L1) cases. Below these are the corresponding variation of the \emph{Fourier transforms} with $k$ --- the deviation from the $K$-point, for the $B\neq0$ (R2 -- see \eqref{eq-FTparam} for parameters) and $B=0$ (L2) cases. All four plots were made using Mathematica\cite{2008-mathematica-yq}.}
\label{fig-results}
\end{figure*}

In this work we have followed the lattice-related conventions used in \cite{2007-manes-yq}. The hamiltonian near the $K^{*}$-point is related to that near the $K$-point by a parity transformation of the lattice: $\mc{H}(-K+k) = \s_{x}\mc{H}(K-k)\s_{x}$. Since the low energy theory describing free excitations near the $K/K^{*}$ points obey a (2+1)-dimensional Dirac theory\cite{1947-wallace-yq,1956-mcclure-dn}, we can again use the parity operator $\s_{z}$ for the dirac fields within a given valley to relate the stationary eigenstates of and the contributions to the propagator/Green's function from the two valleys (\emph{modulo} the $e^{\pm i\mbs{K}.\mbs{r}}$/$e^{\pm 2i\mbs{K}.\mbs{r}}$ factors):
\begin{align}\label{eq-relatevalleys}
\c^{-\mbs{K}}_{\mbs{k},s} &\propto i\s_{y} \c_{\mbs{k},s}^{\mbs{K}}\nn\\
\mb{G}_{-\mbs{K}} &= \s_{y}\mb{G}_{\mbs{K}}\s_{y}^{\dag} \equiv -\s_{y}\mb{G}_{\mbs{K}}\s_{y}
\end{align}
The total low-energy electronic green's function will finally be given by:
\begin{align}\label{eq-totalG}
\mb{G} = \mb{G}_{\mbs{K}} + \mb{G}_{-\mbs{K}}
\end{align}
When a finite perpendicular magnetic field $B \hat{z}$ is present, in a convenient gauge choice the energy eigenstates at the $K$-point are given by (for $B>0$):
\begin{align}\label{eq-states2}
&\vx_{n, k}^{\mbs{K}}(\mbs{r}) = \frac{e^{ikx}e^{i \mbs{K}.\mbs{r}}}{\sqrt{\g_{n} L_{x}}}\le(\ba{c} - \s_{n}\f_{|n|-1}(y-k\ell^{2})\\ \f_{|n|}(y-k\ell^{2})\ea\ri)
\end{align}
where $\s_{n} = \sgn(n)(1-\d_{n0})$, $\ell = \sqrt{\hbar/(eB)}$, $\g_{n} = 2 - \d_{n,0}$, $\w_{c} = \sqrt{2} v/\ell$, $E_{n, k} = \s_{n} \sqrt{|n|} \hbar \w_{c}$ and $n\in \mb{Z}$. The $\f_{n}$'s are the orthonormal eigenfunctions of the 1D Simple Harmonic Oscillator\footnote{$\f_{n}(x) = \frac{1}{\sqrt{2^{n}n!\sqrt{\p}\ell}}e^{-\frac{x^{2}}{2\ell^{2}}} H_{n}(x/\ell)$ where $H$ are the Hermite polynomials.} (take $\f_{-1}\equiv0$). The $k$'s are consistent with periodic boundary conditions in the $x$-direction.Since the SHO eigenfunctions need to be confined inside the sample, we end up with a degeneracy of $\mc{N}=L_{x}L_{y}/(2\p \ell^{2})$ per Landau level.

The green's function may be calculated as:
\begin{align}\label{eq-greens2}
&\mb{G}_{\mbs{K}}(\mbs{r'}, \mbs{r}, z) = \sum_{n,k}\frac{\vx_{n, k}^{\mbs{K}}(\mbs{r'})\vx^{\mbs{K}\,\dag}_{n, k}(\mbs{r})}{z-E_{n, k}}\nn\\
&= \frac{e^{i \mbs{K}.\mbs{\varrho}}}{2\p\ell^{2}}\sum_{n}\frac{e^{-\frac{\rr^{2}}{4\ell^{2}} - i\x(y+y')/2}}{\g_{n}(z-E_{n})}\times\nn\\
&\le(\ba{cc} \s_{n}^{2} L_{|n|-1} & i\s_{n}\frac{\rr}{\sqrt{2|n|}\ell} e^{-i\th}L^{1}_{|n|-1} \\ i\s_{n}\frac{\rr}{\sqrt{2|n|}\ell} e^{i\th}L^{1}_{|n|-1} & L_{|n|} \ea\ri)\\
&\equiv \frac{e^{i \mbs{K}.\mbs{\varrho}}}{2\p\ell^{2}}\sum_{n}\frac{e^{-\frac{\rr^{2}}{4\ell^{2}} - i\x(y+y')/2}}{\g_{n}(z-E_{n})}\mb{M}^{\mbs{K}}_{n}(\mbs{\varrho})\nn\\
&\equiv e^{- i\x(y+y')/2} \mb{N}^{\mbs{K}}(\mbs{\varrho}, z)\nn
\end{align}
where $\rr$ and $\th$ are the modulus and the argument respectively of the complex number $\varrho = (x'-x) + i(y'-y)$ and $\mbs{\varrho} = (x'-x,y'-y)$; the argument of the (associated) Laguerre polynomials, denoted by $L$, is $\frac{\rr^{2}}{2\ell^{2}}$ and we have defined the matrices $\mb{M}$ and $\mb{N}$ for later reference.

We note here that we haven't found this convenient elementary result in the literature till date.

The $K^{*}$ point eigenfunctions as well as the Green's functions are related to those at the $K$ point by the aforesaid relations \eqref{eq-relatevalleys}.

It is instructive to compare the zero field and finite field free electronic green's functions. A comparative plot of the angular average of one of their components is shown in Figure \ref{fig-comparegfs}, plotted against the spatial separation as a fraction of the cyclotron diameter $d_{cyc} = 2|E|/(evB) = 2\sqrt{2n_{LL}}\ell_{B}$.

We now consider the case when an impurity potential $\mb{V}(\mbs{r})$ (whose spatial fourier transform is given by $\tilde{\mb{V}}(\mbs{q})$) is present in graphene. We consider the general form of $\mb{V}(\mbs{r})$ in what follows --- i.e, $\mb{V}(\mbs{r})$ is a \emph{general hermitian 2$\times$2 matrix function} of $\mbs{r}$ ($\mb{V}(\mbs{r}) = \mb{V}^{\dag}(\mbs{r})$).

Scanning Tunneling Microscopes (STMs) give a signal corresponding to the local value of the spectral function \cite{2004-bruus-lq}. We can use the Green's function derived above to obtain the spatial FT of the change in the LDOS (given by the change in the spectral function) to the linear order in the impurity potential strength as follows:
\begin{align}\label{eq-spectral1}
&\d A(\mbs{r}, \w)\nn\\
&= - 2\Im\tr\le[\int d\mbs{r'}\mb{G}(\mbs{r}, \mbs{r'}, \w+i\y)\mb{V}(\mbs{r'})\mb{G}(\mbs{r'}, \mbs{r}, \w+i\y)\ri]\nn\\
&\imply \d \tilde{A}(\mbs{k}, \w) = - 2\Im\tr\le[\tilde{\mb{X}}(\mbs{k}, \w+i\y)\tilde{\mb{V}}(\mbs{k})\ri]
\end{align}
In the above, $\tilde{\mb{X}}(\mbs{k}, z)$ is the fourier transform (w.r.t $\mbs{\varrho}$) of 
\begin{align}\label{eq-spectralpoints}
&\mb{X}(\mbs{\varrho}, z) = \mb{N}(-\mbs{\varrho}, z)\mb{N}(\mbs{\varrho}, z)\nn\\
&= \mb{N}^{\mbs{K}}(-\mbs{\varrho}, z)\mb{N}^{\mbs{K}}(\mbs{\varrho}, z) + \mb{N}^{-\mbs{K}}(-\mbs{\varrho}, z)\mb{N}^{-\mbs{K}}(\mbs{\varrho}, z) \nn\\& \;\;\;\;\;+ \mb{N}^{\mbs{K}}(-\mbs{\varrho}, z)\mb{N}^{-\mbs{K}}(\mbs{\varrho}, z)+ \mb{N}^{-\mbs{K}}(-\mbs{\varrho}, z)\mb{N}^{\mbs{K}}(\mbs{\varrho}, z)
\end{align}
From \eqref{eq-states2} and \eqref{eq-greens2} we see that $\mb{N}^{\pm\mbs{K}}(\mbs{\varrho}, z)$ possess the prefactors $e^{\pm i \mbs{K}.\mbs{\varrho}}$. We thus deduce that scattering by the impurity potential $\mb{V}$ will yield spatial LDOS oscillations around the wavevectors $\mbs{0}$ (terms in the first row of \eqref{eq-spectralpoints}) and $\pm2\mbs{K}$ (second row of \eqref{eq-spectralpoints}) as well as those joined to these by reciprocal lattice vectors, due to intra and intervalley scattering respectively. The resulting lattice is identical to the reciprocal lattice (see Figure~\ref{fig-results}(C)).

Since scattering around the zero wavevector can also arise from many slowly varying unknown environmental potentials, we expect that LDOS oscillations around the wavevectors $\pm 2\mbs{K}$ near isolated atomically sharp defects will better reproduce the LDOS profiles that our theory predicts and for this reason we'll focus on explaining how to calculate the features around $\pm2\mbs{K}$. To do this we need to isolate in \eqref{eq-spectral1} the part due to intervalley scattering, which amounts to using the terms in the last line of \eqref{eq-spectralpoints} that we shall refer to as $e^{\mp2 i \mbs{K}.\mbs{\varrho}}\mb{X}_{\mp}$ respectively.

The Landau levels have been assumed to be sharp in the treatment so far and so direct evaluation of  \eqref{eq-greens2} and the subsequent calculations will yield a sum of delta functions in energy. To be able to resolve the spatial functional forms and to reflect realistic experimental conditions we can either assume that the Landau levels are broadened or that the STM has a finite detection window. We have chosen to take a gaussian detection window with width $\G$\footnote{If the Landau levels are sharper than the lock-in AC voltage amplitude applied to the STM, this approach is more appropriate. It is easy to calculate the case of broadened Landau levels (or non-gaussian profiles) and does not affect our general conclusions.}:
\begin{align}\label{eq-spectralobs}
\d \tilde{A}(\mbs{k}, \w)_{\sub{obs}} &\propto \int d\w' \frac{e^{-\frac{(\w' - \w)^{2}}{2\G^{2}}}}{\sqrt{2\p}\G}\d \tilde{A}(\mbs{k}, \w')
\end{align}
It is now possible to write down the LDOS as a series expansion in $\frac{\G}{\D E}$, where $\D E$ is of the order of the difference between the energy levels that are incorporated into the calculation. To see this, we note that upon substituting the expression \eqref{eq-greens2} of the green's function we come across sums of the following structure ($g$ represents the gaussian in \eqref{eq-spectralobs}; $f_{mn}$ is proportional to Fourier transforms of the form $\int d^{2}q\tr[\tilde{\mb{M}}^{\pm\mbs{K}}_{m}(\mbs{q} - \mbs{k})\tilde{\mb{M}}^{\mp\mbs{K}}_{n}(\mbs{q})\tilde{\mb{V}}(\mp 2\mbs{K} + \mbs{k})]$ that satisfy the condition $\Im f_{mn}=-\Im f_{nm}$ when $\mb{V}$ is \emph{invariant under spatial inversion}):
\begin{widetext}
\begin{align}\label{eq-sumsimplify1}
\Im \int d\w' g(\w'-\w)\sum_{m,n}\frac{f_{mn}}{(\w'-\e_{m}+i\y)(\w'-\e_{n}+i\y)}& \!\!\!\!\!\!\!\!\!\!\!\!\!\!\stackrel{\Im f_{mn}=-\Im f_{nm}}{=} \p \sum_{m\neq n}\Re f_{mn}\frac{g(\e_{n}-\w)-g(\e_{m}-\w)}{\e_{m}-\e_{n}} - 2\p \sum_{n}f_{nn}g'(\e_{n} - \w) \nn\\
&= \p \sum_{n}g(\e_{n}-\w)\sum_{m(\neq n)} \frac{\Re (f_{mn} + f_{nm})}{\e_{m} - \e_{n}}- 2\p \sum_{n}f_{nn}g'(\e_{n} - \w)
\end{align}
\end{widetext}
The `diagonal' term involving $f_{nn}$ above (that corresponds to the particle ejected by the STM tip remaining in the same Landau level on both legs of its journey before and after scattering off the defect) gives the main contribution and the other members in the sum are supressed by the aforesaid factors of $\frac{\G}{\D E}$. The numerical calculations that we subsequently perform are taking only the first few terms of this series into account and work well for large magnetic fields when $\D E\propto \sqrt{B} \gg \G$.

From \eqref{eq-spectral1} and subsequent discussions we find that we can write $\d \tilde{A}_{\sub{obs}}$ in the form $\d \tilde{A}(\pm 2\mbs{K} + \mbs{k}, \w)_{\sub{obs}} = \tr[\tilde{\mb{D}}_{\pm}(\mbs{k}, \w)\tilde{\mb{V}}(\pm 2\mbs{K} + \mbs{k})]$, where $\mb{D}_{\pm} = i(\mb{X}_{\pm}-\mb{X}^{\dag}_{\pm})$. We can make the following general comments regarding the functional dependance of the components of $\tilde{\mb{D}}_{\pm}(\mbs{k}, \w)$ as a function of $\mbs{k}$. Let $n_{LL}(\w)$ denote the Landau level index corresponding to the Landau level nearest to the energy of observation $E_{\sub{obs}} = \hbar\w$ --- it is thus the integer closest to $\sgn\w(\w/\w_{c})^{2}$). The `diagonal' term in \eqref{eq-sumsimplify1} that is the most important contribution then corresponds to $n = n_{LL}(\w)$. From the definition of $\mb{D}$ and using \eqref{eq-greens2} we see that $\mb{D}$ consists of products of two oscillatory functions (like those shown in Figure \ref{fig-comparegfs}). We thus expect spatial oscillation scales set by the wavevectors $2\p/(2\sqrt{2\le|n_{LL}\ri|}\ell)$ and twice of $|\w|/v$ to appear in $\mb{D}(\mbs{r})$. Our calculation confirms this expectation --- we find that $\tilde{\mb{D}}(\mbs{k})$ displays a set of about $\le|n_{LL}\ri|$ oscillatory peaks starting at $k=0$ and separated by a period $\D k$ (see below); it then decays rapidly after a maximum wavevector $k_{\sub{max}}$, where
\begin{align}\label{eq-FTparam}
\D k\sim \frac{2}{\ell}\sqrt{\frac{2}{\le|n_{LL}\ri|}} = \frac{8}{d_{\sub{cyc}}},\, k_{\sub{max}} = \frac{2|\w|}{v} \sim 2\sqrt{2}\frac{\le|n_{LL}\ri|}{\ell}
\end{align}

The off-diagonal elements in \eqref{eq-greens2} possess an angular dependence and for this reason $\tilde{\mb{D}}(\pm 2 \mbs{K} + \mbs{k})$ exhibits sinusoidal oscillations in $\th_{k}$ and $2\th_{k}$ for a given $k$, $\th_{k}$ being the orientation angle of $\mbs{k}$ with respect to the direction of $\mbs{K}$. We find that when intravalley scattering is considered, only the off-diagonal components of $\mb{V}$ give rise to $\th_{k}$-oscillations while in the case of intervalley scattering, the diagonal components of $\mb{V}$ can, in addition, lead to $2\th_{k}$-oscillations.

The results of our calculations have been summarized in the Figure \ref{fig-results}. The FT of the LDOS oscillations is plotted near a short-ranged diagonal impurity potential $\tilde{\mb{V}}(\mbs{k}) \propto \mb{I}$. Given any other nontrivial form of this potential, the LDOS modulations may be found straightforwardly from the above prescription.

It is worth noting here that we have only quantified the oscillation parameters that may be observed in the spatial Fourier transform and \emph{not the Power Spectrum}, examples of which are however also shown in Figure \ref{fig-results} (L1, R1). Upon squaring the FT modulus to obtain the power spectrum the result could have twice as many oscillations -- this needs to be kept in mind when comparing the foregoing results with experimental signatures.

In conclusion, in this work we have laid out the framework for calculating the LDOS modifications around an impurity in graphene in the presence of a strong magnetic field. We  use the linearly dispersing chiral quasiparticle theory. To calculate the QPI we have derived the graphene green's function in a magnetic field. There are two distinct regimes -- in case of a strong field we have a situation of QHE while in the opposite case of a weak field the level broadening $\Gamma$ (due to lock-in modulation of STM voltage or due to impurity scattering, etc) will be larger then the Landau level splitting $\Delta E \sim B^{1/2}$. We considered the case of a strong magnetic field. To this end we established a series expansion in $\Gamma/\Delta E$. In this limit our approach can be used to obtain the LDOS oscillations for any impurity potential. While the exact form of these oscillations vary by impurity type, we have identified a few important characteristics that may be observed in the FT of these oscillations --- impurity-induced LDOS modulations in a magnetic field thus offers an alternative avenue for Landau level spectroscopy using local probes.\\
(Note added: We recently became aware of the preprint \cite{2009-bena-yq} where similar questions have been addressed.)

We acknowledge useful discussions with V.\ Brar, M.\ Crommie, H.\ Dahal, B.\ I.\ Halperin, J.\ Lau, S.\ Sachdev, N.\ C.\ Yeh,  T.\ Wehling and  Y.\ Zhang. We are particularly grateful to H.\ Manoharan and L.\ Mattos for useful discussions and for sharing their preliminary STM data with us. This work was supported by the US DOE through LDRD and BES and the University of California UCOP-09-027 funds at LANL. RRB would also like to acknowledge support from the Harvard University Physics Department.

\bibliography{qpi}

\end{document}